\documentclass[10pt,american,preprint, aps, nofootinbib]{revtex4-1}
\usepackage[T1]{fontenc}
\usepackage[latin9]{inputenc}
\usepackage{babel}
\usepackage{mathrsfs}
\usepackage{amsmath}
\usepackage{amssymb}
\usepackage[pdfusetitle,
 bookmarks=true,bookmarksnumbered=false,bookmarksopen=false,
 breaklinks=false,pdfborder={0 0 1},backref=false,colorlinks=false]
 {hyperref}

\makeatletter

\providecommand{\tabularnewline}{\\}

\usepackage{comment}
\usepackage{etoolbox}
\usepackage{breqn}

\makeatletter
\let\cat@comma@active\@empty
\makeatother

\newcommand{\breqnoverloadothers}
{%
    \renewenvironment{equation}{\ignorespaces\begin{dmath}}{\end{dmath}\ignorespacesafterend}%
    \renewenvironment{equation*}{\ignorespaces\begin{dmath*}}{\end{dmath*}\ignorespacesafterend}%
    \renewenvironment{multline}{\ignorespaces\begin{dmath}}{\end{dmath}\ignorespacesafterend}%
    \renewenvironment{multline*}{\ignorespaces\begin{dmath*}}{\end{dmath*}\ignorespacesafterend}%

}

\newcommand\breqnundefineothers
{%
    \renewenvironment{equation}{}{}%
    \renewenvironment{equation*}{}{}%
    \renewenvironment{multline*}{}{}%

}

\AtBeginEnvironment{dmath}{\breqnundefineothers}
\AtBeginEnvironment{dmath*}{\breqnundefineothers}

\AtBeginEnvironment{dgroup}{\def\breqnundefineothers{}\breqnoverloadothers}
\AtBeginEnvironment{dgroup*}{\def\breqnundefineothers{}\breqnoverloadothers}

\newcommand\brwrap[3]{%
  \setbox0=\hbox{$#2$}
  \left#1\vbox to \the\ht0{\hbox to 0pt{}}\right.\kern-.2em
  \begingroup #2\endgroup\kern-.15em
  \left.\vbox to \the\ht0{\hbox to 0pt{}}\right#3
}

\makeatother

\begin{document}
\title{Generalized Effective Field Theory for Four-Dimensional Black Hole
Evaporation}
\author{Bing-Nan Liu}
\author{David A. Lowe}
\affiliation{Department of Physics, Brown University, Providence, RI 02912, USA}
\author{Larus Thorlacius}
\affiliation{Science Institute, University of Iceland, Dunhaga 3, 107 Reykjavik,
Iceland}
\begin{abstract}
The quantum induced stress tensor of 3+1-dimensional Einstein gravity,
with conformally coupled matter, is studied in an effective field
theory approach. In this context, Riegert's non-local effective action
is sufficient to reproduce the trace anomaly in curved spacetime but
in general the effective action can include additional non-local but
scale invariant terms that influence the semiclassical physics without
affecting the trace anomaly. Here, a truncated model, with only one
additional term involving the square of the Weyl tensor, is used to
find the induced stress tensor in a black hole background. With suitable
physical conditions, a solution of the resulting 4th order equations
leads, in a static limit, to a unique quantum state matching expected
properties of the Unruh state.
\end{abstract}
\maketitle

\section{Introduction}

The semiclassical dynamics of black holes in 3+1 dimensions, taking
into account the back-reaction on the black hole metric due to quantum
mechanical emission of Hawking radiation, remains a challenging theoretical
problem that should in principle be amenable to effective field theory
methods. A promising step in this direction, is to consider Riegert\textquoteright s
covariant, non-local action \citep{RIEGERT198456}, which can be expanded
around flat spacetime and reproduces the trace anomaly in a curved
spacetime background. In recent work \citep{Lowe:2025bxl}, Riegert's
action was studied in a static black hole background and the quantum
induced stress energy tensor was computed analytically at the semiclassical
level. These calculations confirmed earlier work of \citep{Mottola:2006ew,Mottola:2016aa,Mottola:2025fhl}
but reinterpreted the boundary conditions of the resulting higher
derivative theory of gravity. The main result of \citep{Lowe:2025bxl}
was that for smooth initial data along a Cauchy surface, restricting
to a time-independent stress tensor, a unique solution is obtained
such that the stress tensor is non-singular on the future event horizon.
In particular no quantum hair is observed, implying Hawking radiation
is generated in a deterministic way at the semiclassical level, despite
the non-locality of the theory. 

The unique solution for the induced stress tensor has, however, the
serious drawback that the outgoing energy flux of Hawking radiation
is \emph{negative} for conformally coupled matter fields with spin
$\leq1$. In \citep{Lowe:2025bxl} this was remedied by artificially
imposing a positive sign on the anomaly coefficient denoted by $b$
below, which leads to a positive outgoing energy flux. Alternatively,
the matter families considered in \citep{Lowe:2025bxl} could be thought
of as some exotic massless higher spin conformally coupled fields
whose origin was not studied. However, given its simplicity versus
more general approaches such as in \citep{Lowe:2022tun}, it will
be tempting to use the unadorned Riegert model (with the artificial
positive sign for $b$) for future exploration of dynamical black
hole solutions with back-reaction included.

In the present work we consider a generalization of Riegert's action
by including an additional non-local term involving two powers of
the square of the Weyl tensor. This action has been previously studied
in \citep{Balbinot:1999ri,Mazur2001,Mottola:2006ew,Anderson:2007aa}.
With the extra term in the effective action in place we can relax
the artificial constraint on the anomaly coefficient and work with
conventional matter fields. We proceed to localize the extended non-local
action by introducing two scalar fields which satisfy fourth order
equations of motion. The semiclassical induced stress tensor is derived
and computed in a Schwarzschild black hole background in analytic
form. The conclusions are similar to \citep{Lowe:2025bxl} in that
for smooth initial data, restricting to a time-independent stress
tensor, the outgoing flux is uniquely determined. This time around
the outgoing flux is a sum of two terms: the negative contribution
from the original Riegert action and a positive definite contribution
that depends on a free parameter in the action, not predicted by the
anomaly coefficients. The free parameter can be fixed by matching
to the known luminosity of Hawking radiation. 

Our final results largely reproduce the expected asymptotic form,
noted in \citep{Christensen:1977jc}, of the quantum stress tensor
near the future horizon and at future null infinity. A relatively
minor discrepancy is a logarithmic enhancement in the leading falloff
behavior of the transverse components of the stress tensor at future
null infinity compared to \citep{Christensen:1977jc}. The results
of \citep{Barvinsky:2023exr} suggest the exact one-loop gravitational
action can be interpreted as the Riegert term together with an infinite
series of conformally invariant non-local terms. The model presented
here, corresponds to a truncation of this infinite sequence to a finite
number of terms based on the simplest known conformal invariants.
The asymptotic behavior of the stress tensor may be further improved
by including additional conformally invariant terms, but at the price
of significantly complicating the formalism.

\section{Generalized Riegert Model\label{sec:Generalized-Riegert-Model}}

Our starting point is the following general expression for the leading
order trace anomaly of the stress tensor in curved 3+1-dimensional
spacetime, for classically conformally coupled fields,
\begin{equation}
g^{ab}\left\langle T_{ab}\right\rangle =\frac{1}{16\pi^{2}}\left(aC^{2}+bE-c\nabla^{2}R+dR^{2}+eF^{2}\right)\,,\label{eq:traceanom}
\end{equation}
where $a,b,c,d,e$ are coefficients that depend on the matter content
of the theory\footnote{We adopt the notation of Riegert's original paper \citep{RIEGERT198456}
for the anomaly coefficients, but we use Misner-Thorne and Wheeler's
$(+++)$ conventions \citep{misner} throughout, while Riegert uses
Birrell and Davies conventions \citep{Birrell:1982ix}. As a result,
some signs are reversed, including in the $c\nabla^{2}R$ term in
\eqref{eq:traceanom}.} and
\begin{align}
C^{2} & =R^{abcd}R_{abcd}-2R^{ab}R_{ab}+\frac{1}{3}R^{2},\label{eq:c2e}\\
E & =R^{abcd}R_{abcd}-4R^{ab}R_{ab}+R^{2},\nonumber 
\end{align}
are the square of the Weyl tensor and the Euler density, respectively.
The constant $e$ is proportional to the beta function of the gauge
theory. In the present work we do not include gauge fields so set
$e=0$. Riegert \citep{RIEGERT198456} imposed $d=0$ as a condition
on the collection of matter fields in the theory but Duff has shown
this condition holds for ordinary conformally coupled matter fields
with spin $\leq1$ \citep{Duff:1977ay,Duff:1993wm}. 

The anomaly coefficients, in terms of the number of matter species
with spins $\leq1$, are given by
\begin{align}
a & =\frac{1}{120}\left(n_{s}+6n_{f}+12n_{V}\right),\nonumber \\
b & =-\frac{1}{360}\left(n_{s}+11n_{f}+62n_{V}\right),\label{eq:abcd}\\
c & =-\frac{1}{180}\left(n_{s}+6n_{f}+12n_{V}\right),\nonumber \\
d & =0\,,\nonumber 
\end{align}
where $n_{s},n_{f}$ and $n_{V}$ are the number of scalars, Dirac
fermions and vectors respectively \citep{Duff:1993wm}. In order for
the semiclassical approximation to be under control the non-vanishing
coefficients are taken to be of order $N\gg1$ (while scaling $\hbar\sim1/N$)
so that the matter field contribution to the effective action is dominant
compared to that of the metric sector. 

We note that the $b$ and $c$ coefficients in \eqref{eq:abcd} are
negative for any combination of low spin fields. As it turns out,
the outgoing semiclassical energy flux from a black hole is independent
of $c$ but it does depend on $b$. In particular, in our earlier
work on black hole emission in the Riegert model \citep{Lowe:2025bxl},
we found the sign of the outgoing energy flux to be determined by
the sign of $b$. This led us to impose the condition $b>0$, which
does not hold for ordinary spin $\leq1$ fields. In the present work,
we show how this this restriction can be avoided by introducing additional
scale invariant terms in the effective action.

The trace anomaly (with $d=e=0$) is reproduced by the following scalar-tensor
theory, with two auxiliary scalar fields, which was previously studied
in \citep{Balbinot:aa,Mazur2001,Mottola:2006ew,Anderson:2007aa},

\begin{dmath}
\begin{equation}
S=\int d^{4}x\,(-g)^{1/2}\left[\frac{1}{16\pi}R+\frac{1}{192\pi^{2}}(c-\tfrac{2}{3}b)R^{2}-\tfrac{b}{2}\nabla^{2}\phi\nabla^{2}\phi-\tfrac{b}{3}R(\nabla\phi)^{2}+bR^{ab}\nabla_{a}\phi\nabla_{b}\phi+\frac{\phi}{8\pi}\left((a+b)C^{2}+\frac{2b}{3}\left(R^{2}-3R_{ab}R^{ab}-\nabla^{2}R\right)\right)-\tfrac{1}{2}\nabla^{2}\chi\nabla^{2}\chi-\tfrac{1}{3}R(\nabla\chi)^{2}+R^{ab}\nabla_{a}\chi\nabla_{b}\chi+fC^{2}\chi\right].\label{eq:action}
\end{equation}
\end{dmath} The terms involving the $\chi$ field do not contribute
to the anomaly but they are crucial to producing physically sensible
results for low-spin matter fields, as we will see below. 

The metric equation of motion is given by a somewhat lengthy expression
to be found in the Appendix. The result was obtained with the aid
of the symbolic algebra package xAct \citep{Martin-Garcia:2007bqa,Martin-Garcia:2008ysv,Brizuela:2008ra}.

The scalar field equations are fourth order in derivatives, but linear
in the scalars,

\begin{dmath}
\begin{equation}
0=-\frac{(a+2b)R_{ab}R^{ab}}{4\pi}+\frac{(a+3b)R^{2}}{24\pi}+\frac{(a+b)R_{abcd}R^{abcd}}{8\pi}-\frac{b\nabla^{2}R}{12\pi}+\tfrac{2}{3}bR\nabla^{2}\phi-\tfrac{1}{3}b\nabla^{a}R\nabla_{a}\phi-2bR^{ab}\nabla_{b}\nabla_{a}\phi-b\nabla^{2}\nabla^{2}\phi\,,\label{eq:scalareom}
\end{equation}
\end{dmath}and
\begin{equation}
0=-2fR_{ab}R^{ab}+\tfrac{1}{3}fR^{2}+fR_{abcd}R^{abcd}+\tfrac{2}{3}R\nabla^{2}\chi-\tfrac{1}{3}\nabla^{a}R\nabla_{a}\chi-2R^{ab}\nabla_{a}\nabla_{b}\chi-\nabla^{2}\nabla^{2}\chi\,.\label{eq:scalareom2}
\end{equation}

As a simple check on these expressions, one can take the trace of
the induced stress tensor listed in \eqref{eq:stress-g}-\eqref{eq:stress}
in the Appendix,
\begin{equation}
g^{ab}\left\langle T_{ab}\right\rangle =\frac{b}{2\pi}\left(\nabla^{2}\nabla^{2}\phi-\tfrac{2}{3}R\nabla^{2}\phi+2R^{ab}\nabla_{a}\nabla_{b}\phi+\tfrac{1}{3}\nabla^{a}R\:\nabla_{a}\phi\right)-\frac{1}{16\pi^{2}}\left(c-\frac{2}{3}b\right)\nabla^{2}R\,,\label{eq:stresstrace}
\end{equation}
and eliminate the scalar fields using their equations of motion \eqref{eq:scalareom},
\eqref{eq:scalareom2}. After some straightforward algebra one then
recovers the trace anomaly formula \eqref{eq:traceanom} with $d=e=0$. 

\section{Non-Local Gravitational Action}

If we integrate out the two scalar fields, we obtain a non-local gravitational
action of the form \begin{dmath}
\begin{equation}
S=\int d^{4}x\,(-g)^{1/2}\left[R+f^{2}C^{2}\frac{1}{\varDelta^{4}}C^{2}+\left((a+b)C^{2}+\frac{2b}{3}\left(R^{2}-3R_{ab}R^{ab}-\nabla^{2}R\right)\right)\frac{1}{b\varDelta^{4}}\left((a+b)C^{2}+\frac{2b}{3}\left(R^{2}-3R_{ab}R^{ab}-\nabla^{2}R\right)\right)\right],\label{eq:nonlocalaction}
\end{equation}
\end{dmath}where 
\begin{equation}
\varDelta^{4}=\left(\nabla^{2}\right)^{2}+2R^{\mu\nu}\nabla_{\mu}\nabla_{\nu}-\frac{2}{3}R\nabla^{2}+\frac{1}{3}\left(\nabla^{\mu}R\right)\nabla_{\mu}\label{eq:d4op}
\end{equation}
is the unique conformally covariant 4th order operator acting on a
scalar of vanishing scale dimension \citep{RIEGERT198456} (rewriting
in the conventions of the present paper). The trace anomaly does not
uniquely determine the gravitational effective action and in \eqref{eq:nonlocalaction}
we have included a conformally invariant $C^{2}\frac{1}{\varDelta^{4}}C^{2}$
term. This choice is by no means unique. In fact, an infinite sequence
of conformally invariant non-local terms is expected to arise, already
at one-loop order \citep{Barvinsky:2023exr}. Provided this sequence
of conformally invariant terms behaves as an asymptotic series, one
may hope to capture key aspects of the physics by truncating after
a finite number of leading terms. The truncation in the present paper
seems to be a self-consistent model, with the attractive feature that
it is linearly stable around flat spacetime, though we lose the ability
to justify it as an exact integration out of $N$ spin $\leq1$ matter
fields. The parameters of the truncated model can be adjusted to obtain
the luminosity of Hawking radiation from a black hole but beyond that
the model is not expected to precisely match results derived from
an exact treatment. The full semiclassical equations of motion for
$N$ scalars are set up in \citep{Lowe:2022tun} and these do not
reduce to such a simple local 4th order form in any obvious way. Alternatively,
an effective action derived via heat kernel methods can be found in
\citep{barvinsky1994,barvinsky2009covariantperturbationtheoryiv,Barvinsky:2023exr}
at third order in the curvature (sufficient to reproduce the one-loop
trace anomaly) which in principle should give compatible answers,
though is similarly unwieldy. 

\section{Scalar Field Solutions}

The scalar equations of motion \eqref{eq:scalareom} and \eqref{eq:scalareom2}
can be explicitly solved on the Schwarzschild black hole background,
\begin{equation}
ds^{2}=-\left(1-\frac{2M}{r}\right)dt^{2}+\frac{dr^{2}}{1-\frac{2M}{r}}+r^{2}d\Omega^{2}.\label{eq:schwarzschild}
\end{equation}
Since $R_{ab}=R=0$ in this background, the scalar equations reduce
to 
\begin{align}
\nabla^{2}\nabla^{2}\phi & =\frac{(a+b)R_{abcd}R^{abcd}}{8\pi b}\,,\label{eq:reducedeom}\\
\nabla^{2}\nabla^{2}\chi & =fR_{abcd}R^{abcd}\,.\nonumber 
\end{align}
The two equations only differ by the strength of the Kretschmann scalar
source on the right hand side. The general static spherically symmetric
solution for each scalar field, $\phi$ and $\chi$, can be written
as a linear combination of four independent solutions to the corresponding
homogenous problem plus special solutions, $\phi_{P}$ and $\chi_{P}$,
that satisfy the respective inhomogeneous equations,
\begin{align}
\phi & =c_{1}\phi_{1}+c_{2}\phi_{2}+c_{3}\phi_{3}+c_{4}+\phi_{P}\,,\label{eq:gensolution}\\
\chi & =d_{1}\phi_{1}+d_{2}\phi_{2}+d_{3}\phi_{3}+d_{4}+\chi_{P}\,,\nonumber 
\end{align}
where
\begin{align}
\phi_{1} & =\log\left(1-\frac{2M}{r}\right)\,,\nonumber \\
\phi_{2} & =r^{2}+4Mr+8M^{2}\log r\,,\nonumber \\
\phi_{3} & =-\mathrm{Li}_{2}\left(\frac{2M}{r}\right)+\frac{r}{4M}+\frac{3}{2}\log r-\frac{1}{16M^{2}}\left(r^{2}+4Mr-8M^{2}\log r\right)\log\left(1-\frac{2M}{r}\right)\,,\nonumber \\
\phi_{P} & =-\frac{(a+b)}{8\pi b}\psi(r)\,,\nonumber \\
\chi_{P} & =-f\,\psi(r)\,,\nonumber \\
\psi(r) & =\frac{2}{3M}r+2\log r+\frac{1}{12M^{2}}\left(r^{2}+4Mr+8M^{2}\log r\right)\log\left(1-\frac{2M}{r}\right)\,.\label{eq:solutions}
\end{align}
The eight constants $c_{i}$ and $d_{i}$ are to be fixed using boundary
conditions. The static solution was obtained previously in a somewhat
different form in \citep{Mottola:2006ew}.

\section{The Induced Stress Tensor\label{sec:Stress-Tensor}}

Our goal is to evaluate the semiclassical stress tensor subject to
suitable boundary conditions applied to the asymptotic form of the
stress tensor as $r\to2M$ and $r\to\infty$. A similar strategy was
carried out in \citep{Balbinot:aa} . The stress tensor we have derived,
which appears in the appendix, differs from the stress tensor in \citep{Balbinot:aa}. 

We consider scalar fields of the form
\begin{align}
\phi(t,r) & =d_{\phi}\,t+\phi(r)\,,\label{eq:scalarsol}\\
\chi(t,r) & =d_{\chi}\,t+\chi(r)\,,\nonumber 
\end{align}
where $\phi(r)$ and $\chi(r)$ are static solutions of the form \eqref{eq:gensolution}
and $d_{\phi}$ and $d_{\chi}$ are two additional free parameters
to be fixed by boundary conditions. This satisfies the full scalar
equations of motion \eqref{eq:scalareom} and \eqref{eq:scalareom2}.
The linear time dependence breaks time-translation invariance and
gives rise to a non-vanishing $T_{rt}$ component, yet all the components
of the stress tensor remain time-independent. 

Many of the terms that make up the stress tensor in \eqref{eq:stress-g}-\eqref{eq:stress}
vanish on a Schwarzschild background, where $R_{ab}=R=0$. Despite
this simplification, we obtain rather lengthy expressions (that we
do not write out explicitly here) when we insert \eqref{eq:scalarsol}
for the auxiliary scalar fields. The stress tensor is independent
of $c_{4}$ and $d_{4}$ so these will remain a free parameters, which
do not affect observable quantities.\footnote{We also note that a shift in $d_{4}$ induces a $C^{2}$ term in the
Lagrangian, which is a possible local, scale invariant contribution.
Therefore we do not include a separate local $C^{2}$ term.} 

We are left with eight independent parameters to be determined. Some
of these parameters are fixed by requiring a freely falling observer
crossing the future horizon see a finite energy density. Some of the
same conditions come from requiring the $T_{\theta}{}^{\theta}$ component
of the stress tensor to be finite at the future horizon, which amounts
to a simpler calculation. Near the horizon,
\begin{equation}
T_{\theta}^{\,\,\theta}=\frac{A_{1}}{(r-2M)^{2}}+\frac{2A_{1}}{M(r-2M)}+A_{2}\log^{2}\left(\frac{r}{2M}-1\right)+A_{3}\log\left(\frac{r}{2M}-1\right)+\cdots\,,\label{eq:tthetatheta}
\end{equation}
while near infinity,
\begin{equation}
T_{\theta}^{\,\theta}=B_{1}+\frac{B_{2}}{r}+\frac{B_{3}}{r^{2}}+\frac{B_{4}}{r^{3}}+\cdots\,.\label{eq:tthth}
\end{equation}
where the $A_{i}$ and $B_{i}$ are quadratic functions of the $c_{i}$
and $d_{i}$. 

The conditions $B_{1}=B_{2}=B_{3}=0$ are solved by $c_{2}=d_{2}=0$.
With $c_{2}\neq0$ and $d_{2}\neq0$ there are additional $\frac{\log r}{r^{3}}$
terms. Finiteness near the horizon requires setting $A_{1}=A_{2}=A_{3}=0$.
The $A_{2}$ coefficient involves a sum of two squares and demanding
$A_{2}=0$ on the space of real parameters leads to two conditions,
\begin{equation}
c_{3}=-\frac{a+b}{6\pi b}\,,\quad d_{3}=-\frac{4f}{3}\,.\label{eq:c3d3}
\end{equation}

Let us define the null geodesic vectors $n_{\pm}^{\mu}=\left(\frac{1}{1-\frac{2M}{r}},\pm1,0,0\right)$
in $(t,r,\theta,\phi)$ coordinates. The ingoing flux at past null
infinity $\mathscr{I}^{-}$ is 
\begin{equation}
\frac{1}{4}n_{+}^{\mu}T_{\mu\nu}n_{+}^{\nu}=\frac{C_{3}}{r^{2}}+\cdots\,.\label{eq:nullcontract}
\end{equation}
To find the analog of the Unruh state, with vanishing ingoing flux
we must set $C_{3}=0$. This fixes
\begin{equation}
d_{\chi}=\frac{(a+b)^{2}}{64\pi^{2}b\,f\,M}-\frac{(a+b)}{8\pi f}d_{\phi}+\frac{f}{M}\,,\label{eq:dpvalue}
\end{equation}

Our next requirement is $B_{4}=0$, which is a new condition not present
in \citep{Lowe:2025bxl}. This ensures a faster than $r^{-3}$ falloff
of the angular components of the stress tensor, as is expected of
the Unruh vacuum \citep{Christensen:1977jc}.\footnote{In contrast, the authors of \citep{Balbinot:aa} reported a $r^{-3}$
falloff for $T_{\theta}^{\,\theta}$.} This yields the two branches
\begin{equation}
d_{\phi}=\frac{(a+b)}{8\pi bM}\pm\frac{1}{M}\left(-\frac{f^{2}\bigl(a^{2}+4ab+b(3b+64f^{2}\pi^{2})\bigr)}{b\bigl((a+b)^{2}+64bf^{2}\pi^{2}\bigr)}\right)^{1/2}\,.\label{eq:dphi}
\end{equation}

With these constraints, we return to the near-horizon limit. Regularity
for a freely falling observer requires that 
\begin{align}
n_{-}^{\mu}T_{\mu\nu}n_{-}^{\nu} & =\frac{A_{4}}{(r-2M)^{3}}+\frac{A_{5}}{(r-2M)^{2}}+\frac{A_{6}}{r-2M}+\frac{A_{7}}{r-2M}\log\left(r-2M\right)+\\
 & A_{8}\log^{2}\left(r-2M\right)+A_{9}\log\left(r-2M\right)+\cdots\,,\nonumber 
\end{align}
be finite on the horizon, where again the coefficients $A_{4},\ldots,A_{9}$
are quadratic functions of $c_{i}$ and $d_{i}$. 

Finally we solve the remaining near horizon conditions, $A_{1}=A_{5}=0,$
to fix $c_{1}$ and $d_{1}$. This again leads to a doubling, with
one branch of solutions given by
\begin{align}
c_{1} & =\frac{(a+b)\bigl(3+2\log(2M)\bigr)}{12\pi b}-2\left(-\frac{f^{2}\bigl(a^{2}+4ab+b(3b+64f^{2}\pi^{2})\bigr)}{b\bigl((a+b)^{2}+64bf^{2}\pi^{2}\bigr)}\right)^{1/2},\nonumber \\
d_{1} & =\tfrac{2}{3}f\bigl(3+2\log(2M)\bigr)-\frac{\Bigl(-bf^{2}\bigl((a+b)^{2}+64bf^{2}\pi^{2}\bigr)\bigl(a^{2}+4ab+b(3b+64f^{2}\pi^{2})\bigr)\Bigr)^{1/2}}{4b(a+b)f\pi}\nonumber \\
 & +\frac{16f\pi}{(a+b)}\left(\frac{-bf^{2}\bigl(a^{2}+4ab+b(3b+64f^{2}\pi^{2})\bigr)}{\bigl((a+b)^{2}+64bf^{2}\pi^{2}\bigr)}\right)^{1/2},\label{eq:branch1}
\end{align}
and the other branch by
\begin{align}
c_{1} & =\frac{(a+b)\bigl(3+2\log(2M)\bigr)}{12\pi b}+2\left(-\frac{f^{2}\bigl(a^{2}+4ab+b(3b+64f^{2}\pi^{2})\bigr)}{b\bigl((a+b)^{2}+64bf^{2}\pi^{2}\bigr)}\right)^{1/2},\nonumber \\
d_{1} & =\tfrac{2}{3}f\bigl(3+2\log(2M)\bigr)+\frac{\Bigl(-bf^{2}\bigl((a+b)^{2}+64bf^{2}\pi^{2}\bigr)\bigl(a^{2}+4ab+b(3b+64f^{2}\pi^{2})\bigr)\Bigr)^{1/2}}{4b(a+b)f\pi}\nonumber \\
 & -\frac{16f\pi}{(a+b)}\left(\frac{-bf^{2}\bigl(a^{2}+4ab+b(3b+64f^{2}\pi^{2})\bigr)}{\bigl((a+b)^{2}+64bf^{2}\pi^{2}\bigr)}\right)^{1/2}.\label{eq:branch2}
\end{align}
When the above solutions for $c_{i}$ and $d_{i}$ are inserted, all
the coefficients $A_{i},B_{i}$ and $C_{i}$ vanish without imposing
any additional conditions.

The different branches we have found indicate a degeneracy between
the couplings of certain modes of the two scalars for the special
case of the Schwarzschild background. Different branches give the
same quantum induced stress-tensor. In particular, they yield a unique
prediction for the outgoing null flux at future null infinity $\mathscr{I}^{+}$
, 
\begin{equation}
\frac{1}{4}n_{-}^{\mu}T_{\mu\nu}n_{-}^{\nu}=\frac{2f^{2}}{M^{2}r^{2}}+\frac{\left(a+b\right)^{2}}{32\pi^{2}M^{2}b\,r^{2}}+\cdots\,,\label{eq:outflux}
\end{equation}
corresponding to an object with finite outgoing luminosity. This analytic
solution for the quantum induced stress-energy tensor in the two-scalar
extension of the Riegert model is the main result of this paper. As
a final step one could fix the coupling constant$f$ in this equation
by matching with the known result for the luminosity of Hawking radiation.

The final result respects the asymptotic conditions described in table
1 of \citep{Christensen:1977jc}, assuming conformally coupled scalar
matter. 
\begin{table}

\begin{tabular}{|c|c|c|c|c|}
\hline 
\multicolumn{3}{|c|}{$r\to\infty$} & \multicolumn{2}{c|}{$r\to2M$}\tabularnewline
\hline 
\hline 
 & $T_{\,r}^{r}$ & $T_{\,\theta}^{\theta}$ & $T_{\,r}^{r}$ & $T_{\,\theta}^{\theta}$\tabularnewline
\hline 
CF & $\frac{1}{r^{2}}$ & $\frac{1}{r^{4}}$ & $-(1-\frac{2M}{r})^{-1}$ & $-1$\tabularnewline
\hline 
LLT & $\frac{1}{r^{2}}$ & $\frac{1}{r^{4}}-\frac{\log r}{r^{4}}$ & $-(1-\frac{2M}{r})^{-1}$ & $-1$\tabularnewline
\hline 
\end{tabular}\caption{Asymptotic behavior of the expectation values. Numerical factors are
omitted. The bottom row indicates the results of the present paper,
while the row above shows the expectations of Christensen and Fulling
\citep{Christensen:1977jc}.}

\end{table}
 In particular, with $f$ fixed as described above, the outgoing flux
at infinity is positive, and matches Hawking's prediction. Likewise,
one has vanishing ingoing flux at past null infinity. However, the
$r^{-4}\log r$ behavior at large $r$ indicates the two scalar model
does not give a perfect match to the expected result, presumably because
there are further scale invariant terms in the effective gravitational
action, which have yet to be included, as noted above.

\section{Discussion}

In this work we have constructed and analyzed a generalized version
of Riegert\textquoteright s non-local effective action for four-dimensional
gravity that includes an additional scale-invariant term quadratic
in the Weyl tensor. This term provides an explicit local representation
of the leading conformally invariant corrections beyond the minimal
Riegert action \citep{RIEGERT198456}, while retaining analytic control
in the semiclassical regime. In contrast to the earlier model of \citep{Lowe:2025bxl},
which required a non-standard sign choice of the anomaly coefficient
$b>0$ and thus implied the presence of exotic conformal matter species,
the present formulation accommodates the physically relevant anomaly
coefficients obtained for conformally coupled fields of spin \ensuremath{\le}
1 \citep{Duff:1977ay,Duff:1993wm}. The resulting two-scalar localization
yields a fourth-order but linearly stable system that captures the
dominant quantum effects of the trace anomaly in a tractable, covariant
form.

The principal result of this analysis is the derivation of a closed-form
expression for the semiclassical stress tensor in a Schwarzschild
background, based on the coupled fourth-order field equations for
the auxiliary scalars. By imposing regularity of the components of
the stress tensor on the future horizon and vanishing incoming flux
at past null infinity, the model predicts a unique, time-independent
configuration corresponding to the analog of the Unruh vacuum \citep{Unruh:1976aa}.
In particular, the outgoing null flux at future null infinity is positive
and finite. The solution also does not have arbitrary integration
constants (\textquotedblleft quantum hair\textquotedblright ), confirming
that the semiclassical dynamics are deterministic once physically
sensible boundary conditions are imposed. This behavior parallels
the findings of \citep{Lowe:2025bxl} but is obtained here without
restricting the matter content to nonstandard anomaly coefficients.

The asymptotic form of the derived stress tensor near null infinity
satisfies the boundary conditions originally identified by Christensen
and Fulling \citep{Christensen:1977jc,Candelas:1980zt} for the Unruh
vacuum, as summarized in Table I of the present work. Quantitative
differences in sub-leading terms likely originate from higher-order
conformal invariants not included in the truncated action. Their inclusion
would refine the asymptotic behavior but is not expected to modify
the qualitative structure of the stress tensor or the uniqueness of
the semiclassical state.

Our considerations in this paper were restricted to static black hole
backgrounds, where the finite luminosity of the Unruh state persists
for all time, leading to an infinite ADM mass. The semiclassical model
is also expected to have fully time-dependent solutions, involving
the formation and evaporation of a spherically symmetric black hole,
with full back-reaction included. Given the complexity of the model,
we do not expect to find analytic solutions describing semiclassical
black hole evolution but we hope to report on numerical solutions
in future work.
\begin{acknowledgments}
Work supported in part by the Icelandic Research Fund grant 228952-053. 
\end{acknowledgments}

\appendix

\section*{Appendix: The induced stress tensor\label{sec:Equation-of-Motion}}

The equation of motion resulting from the variation of the action
\eqref{eq:action} with respect to the metric is 
\begin{equation}
G_{ab}=8\pi(T_{ab}^{(g)}+T_{ab}^{(\phi)}+T_{ab}^{(\chi)})\,,\label{eq:stressterms}
\end{equation}
 where $G_{ab}$ is the Einstein tensor and on the right hand side
we have grouped together terms in the induced stress tensor based
on their scalar field dependence, or lack thereof,
\begin{equation}
T_{ab}^{(g)}=-\frac{(c-\frac{2}{3}b)}{48\pi^{2}}\left((R_{ab}-\frac{1}{4}g_{ab}R)R-\nabla_{a}\nabla_{b}R+g_{ab}\nabla^{2}R\right),\label{eq:stress-g}
\end{equation}

\begin{dmath}
\begin{equation}
T_{ab}^{(\phi)}=2b\nabla^{2}\phi\left(\nabla_{a}\nabla_{b}\phi-\frac{1}{4}g_{ab}\nabla^{2}\phi\right)-b(\nabla^{2}\nabla_{a}\phi)\nabla_{b}\phi-b(\nabla^{2}\nabla_{b}\phi)\nabla_{a}\phi+\frac{b}{2}g_{ab}(\nabla^{2}\nabla^{c}\phi)\nabla_{c}\phi-\frac{4b}{3}\left((\nabla_{c}\nabla_{a}\phi)(\nabla^{c}\nabla_{b}\phi)-\frac{1}{4}g_{ab}(\nabla_{c}\nabla_{d}\phi)(\nabla^{c}\nabla^{d}\phi)\right)+\frac{2b}{3}\nabla^{c}\phi\left(\nabla_{c}\nabla_{a}\nabla_{b}\phi-\frac{1}{4}g_{ab}\nabla_{c}\nabla^{2}\phi\right)+\frac{2b}{3}R\left(\nabla_{a}\phi\nabla_{b}\phi-\frac{1}{4}g_{ab}(\nabla\phi)^{2}\right)+\frac{2b}{3}\left(R_{ab}-\frac{1}{4}g_{ab}R\right)(\nabla\phi)^{2}-\frac{4b}{3}\left(R_{acbd}-\frac{1}{4}g_{ab}R_{cd}\right)\nabla^{c}\phi\nabla^{d}\phi-b(R_{ac}\nabla_{b}\phi+R_{bc}\nabla_{a}\phi-\frac{1}{2}g_{ab}R_{cd}\nabla^{d}\phi)\nabla^{c}\phi+\frac{(a+b)}{\pi}\left(R_{a}{}^{c}R_{bc}-\frac{1}{4}g_{ab}R_{cd}R^{cd}\right)\phi-\frac{(a+b)}{2\pi}\left(R_{adef}R_{b}{}^{def}-\frac{1}{4}g_{ab}R_{cdef}R^{cdef}\right)\phi+\frac{b}{\pi}\left(R_{acbd}-\frac{1}{4}g_{ab}R_{cd}\right)R^{cd}\phi-\frac{(a+3b)}{6\pi}R\left(R_{ab}-\frac{1}{4}g_{ab}R\right)\phi-\frac{a}{2\pi}\left(\nabla^{2}R_{ab}-\frac{1}{4}g_{ab}\nabla^{2}R\right)\phi+\frac{a}{6\pi}\left(\nabla_{a}\nabla_{b}R-\frac{1}{4}g_{ab}\nabla^{2}R\right)\phi-\frac{(a+b)}{6\pi}\left(\frac{1}{2}\nabla_{a}R\nabla_{b}\phi+\frac{1}{2}\nabla_{b}R\nabla_{a}\phi-\frac{1}{4}g_{ab}\nabla^{c}R\nabla_{c}\phi\right)+\frac{(3a+b)}{6\pi}\left(\frac{1}{2}\nabla_{a}R_{bc}+\frac{1}{2}\nabla_{b}R_{ac}-\frac{1}{4}g_{ab}\nabla_{c}R\right)\nabla^{c}\phi-\frac{(6a+b)}{6\pi}\left(\nabla_{c}R_{ab}-\frac{1}{4}g_{ab}\nabla_{c}R\right)\nabla^{c}\phi+\frac{(a+3b)}{6\pi}R\left(\nabla_{a}\nabla_{v}\phi-\frac{1}{4}g_{ab}\nabla^{2}\phi\right)-\frac{(3a+4b)}{3\pi}\left(R_{acbd}-\frac{1}{4}g_{ab}R_{cd}\right)\nabla^{c}\nabla^{d}\phi+\frac{(3a+7b)}{6\pi}\left(R_{ab}-\frac{1}{4}g_{ab}R\right)\nabla^{2}\phi-\frac{(3a+5b)}{3\pi}\left(\frac{1}{2}R_{ac}\nabla_{b}\nabla^{c}\phi+\frac{1}{2}R_{bc}\nabla_{a}\nabla^{c}\phi-\frac{1}{4}g_{ab}R_{cd}\nabla^{c}\nabla^{d}\phi\right)-\frac{b}{6\pi}\left(\nabla^{2}\nabla_{a}\nabla_{b}\phi-g_{ab}(\nabla^{2}\nabla^{2}\phi+\frac{1}{4}\nabla_{c}R\nabla^{c}\phi-\frac{1}{2}R\nabla^{2}\phi+\frac{3}{2}R_{cd}\nabla^{c}\nabla^{d}\phi)\right)\,,\label{eq:stress-1}
\end{equation}
\end{dmath}

\begin{dmath}
\begin{equation}
T_{ab}^{(\chi)}=4f\chi\left(R_{a}{}^{def}R_{bdef}-\frac{1}{4}g_{ab}R^{cdef}R_{cdef}+2R_{a}{}^{c}R_{bc}-\frac{1}{2}g_{ab}R^{cd}R_{cd}-\nabla^{2}R_{ab}+\frac{1}{4}g_{ab}R\right)+\frac{4}{3}f\chi\left(\nabla_{a}\nabla_{b}R-\frac{1}{4}g_{ab}\nabla^{2}R-R(R_{ab}-\frac{1}{4}g_{ab}R)\right)+4f\nabla^{c}\chi\left(\nabla_{a}R_{bc}+\nabla_{b}R_{ac}-\frac{1}{2}g_{ab}\nabla^{d}R_{dc}-2\nabla_{c}(R_{ab}-\frac{1}{4}g_{ab}R)\right)-\frac{2f}{3}\left(\nabla_{a}R\nabla_{b}\chi+\nabla_{b}R\nabla_{a}\chi-\frac{1}{2}g_{ab}\nabla_{c}R\nabla^{c}\chi\right)+\frac{4f}{3}R\left(\nabla_{a}\nabla_{b}\chi-\frac{1}{4}g_{ab}\nabla^{2}\chi\right)-8f\nabla^{c}\nabla^{d}\chi\left(R_{acbd}-\frac{1}{4}g_{ab}R_{cd}\right)+4f\nabla^{2}\chi\left(R_{ab}-\frac{1}{4}g_{ab}R\right)-4f\left(R_{a}{}^{c}\nabla_{b}\nabla_{c}\chi+R_{b}{}^{c}\nabla_{a}\nabla_{c}\chi-\frac{1}{2}g_{ab}R^{cd}\nabla_{c}\nabla_{d}\chi\right)-\frac{4}{3}\nabla^{c}\nabla^{d}\chi\left(R_{acbd}-\frac{1}{4}g_{ab}R_{cd}\right)-\nabla^{c}\chi\left(R_{ac}\nabla_{b}\chi+R_{bc}\nabla_{a}\chi-\frac{1}{2}g_{ab}R_{cd}\nabla^{d}\chi\right)+\frac{2}{3}R\left(\nabla_{a}\chi\nabla_{b}\chi-\frac{1}{4}g_{ab}(\nabla\chi)^{2}\right)+\frac{2}{3}(\nabla\chi)^{2}\left(R_{ab}-\frac{1}{4}g_{ab}R\right)+\frac{2}{3}\nabla^{c}\chi\left(\nabla_{c}\nabla_{a}\nabla_{b}\chi-\frac{1}{4}g_{ab}\nabla_{c}\nabla^{2}\chi\right)-\nabla_{a}\chi\nabla^{2}\nabla_{b}\chi-\nabla_{b}\chi\nabla^{2}\nabla_{a}\chi+\frac{1}{2}g_{ab}\nabla^{c}\chi\nabla^{2}\nabla_{c}\chi-\frac{4}{3}\left(\nabla_{c}\nabla_{a}\chi\nabla^{c}\nabla_{b}\chi-\frac{1}{4}g_{ab}\nabla_{c}\nabla_{d}\chi\nabla^{c}\nabla^{d}\chi\right)+2\nabla^{2}\chi\left(\nabla_{a}\nabla_{b}\chi-\frac{1}{4}g_{ab}\nabla^{2}\chi\right)\,.\label{eq:stress}
\end{equation}
\end{dmath}

\bibliographystyle{utcaps}
\bibliography{riegert}

\end{document}